\documentclass[twocolumn,showpacs,prx]{revtex4}
\newcommand\cut[1]{}
\newcommand\Tr{\mathrm{Tr}\;}
\newcommand\atanh{\mathrm{atanh}}
\newcommand\rmi{\mathrm{i}}

\usepackage{amsmath}
\usepackage{tabularx}
\usepackage{graphicx}
\usepackage{color}
\usepackage{epsfig,graphicx}
\usepackage{dcolumn}
\usepackage{amsmath,amssymb}
\usepackage{amsfonts}
\usepackage{url}

\begin{document}

\title[CVM with LR]{A mean field method with correlations determined by linear response}

\author{Jack Raymond}
\affiliation{Dipartimento di Fisica, Universit\`a La Sapienza, Piazzale Aldo Moro 5, I-00185 Roma, Italy}
\author{Federico Ricci-Tersenghi}
\affiliation{Dipartimento di Fisica, Universit\`a La Sapienza, Piazzale Aldo Moro 5, I-00185 Roma, Italy}
\affiliation{INFN--Sezione di Roma 1, and CNR--IPCF, UOS di Roma} 

\begin{abstract}

We introduce a new mean-field approximation based on the reconciliation of maximum entropy and linear response for correlations in the cluster variation method. Within a general formalism that includes previous mean-field methods, we derive formulas improving upon, e.g., the Bethe approximation and the Sessak-Monasson result at high temperature. Applying the method to direct and inverse Ising problems, we find improvements over standard implementations.

\end{abstract}

\maketitle

\section{Introduction: Cluster variational and region based methods}

The cluster variational method (CVM) is a unifying framework for many approximation methods on graphical models, with variational parameters in correspondence with the marginal probability distributions one is interested in~\cite{An:CVM,Pelizzola:CVM,Wainwright:GME}. Fast and provably convergent methods are known for the minimization of the CVM free energy~\cite{Yuille:cccpalgorithms}, and systematic expansion methods about minima have been shown~\cite{Zhou:RGPF,Chertkov:LC}. Generalizations and convex approximations to CVM have allowed for the development of fast and secure inference methods~\cite{yedidia,Weiss:MELP}. A common practice is to use linear response (LR) to improve the correlation estimates~\cite{Welling:LRA,Kappen:BML,Opper:TAPMprl,Opper:TAPMpre,Montanari:CLC}. Another important application of LR has been in inverse problems, the methods have been applied, e.g., to infer protein folding structures and information processing in the retina~\cite{Morcosa:DCA,Cocco:NC}. 

We develop an extension of the standard method for fixing parameters in CVM, that allows the marginal probabilities to be made consistent with the LR estimates. Our model improves over standard implementations on arbitrary graphs for high temperature. From the Bethe approximation, we recover the Sessak-Monasson expression for correlation estimation~\cite{Sessak:SME} from a variational framework, and with an alternative CVM approximation we improve upon the formula. We apply the method to homogeneous lattice models, and demonstrate improvements with respect to the standard implementation. We also apply the method to the inverse problem of estimating couplings given correlations, demonstrating results superior to the best mean-field methods for a range of temperatures. For brevity we focus only on binary variables (spins), pairwise interactions and three standard region selection rules; but the principle we outline is flexible with respect to these criteria. The framework offers many avenues for improvement: e.g., many of the extensions outlined in introductory comments can be directly incorporated.

A paradigmatic problem in physics is the determination of the thermodynamics and marginal probabilities of a system with $N$ spins $\{\sigma_i=\pm 1\}$ with a Hamiltonian determined by external fields $H_i$, and symmetric pair couplings ($J_{ii}=0$ and $J_{ij}=J_{ji}$)
\begin{equation}
\mathcal{H}(\sigma) = - \sum_i H_i \sigma_i - \sum_{i < j} J_{ij} \sigma_i \sigma_j \label{eq:Hamiltonian}\;.
\end{equation}
The free energy
\begin{equation}
  \beta F(H,J) = - \log \Tr \left[\exp (-\beta \mathcal{H}(\sigma))\right] \label{eq:F}\;,
\end{equation}
is in most cases computational intractable for moderate system sizes and the analytic solution is unknown in the large $N$ limit. $\Tr[\cdot]$ denotes a summation over all spin variables in the expression. The cluster variational method (CVM) offers insight into approximations~\cite{An:CVM,Pelizzola:CVM}: the free energy functional
\begin{equation}
  F_{CVM}(b,J,H) =  E(b,J,H) - \frac{1}{\beta} S(b) \label{eq:Fcvm}
\end{equation}
is the sum of energetic part
\begin{equation}
  E(b,J,H) = - \sum_{i<j} J_{ij} \Tr \left[b_{ij} \sigma_{i}\sigma_j\right]
  - \sum_i H_{i} \Tr \left[b_{i} \sigma_{i}\right]\;, \label{eq:Ecvm}
\end{equation} 
and entropic part
\begin{equation}
  S(b) = - \sum_{R} c_R \Tr \left[b_R\log b_R\right] \;,\label{eq:Scvm}
\end{equation}
where $R$ are subsets of variables, $c_R$ are integer counting/M\"obius numbers, $\sigma_R$ denotes the set of variables $\{\sigma_i : i \in R\}$, and $b_R(\sigma_R)$ are beliefs over the variables in $R$ (arguments omitted for brevity).
Selecting a set of regions that forms a junction tree (e.g.\ selecting all possible regions is sufficient) the approximation is exact, Eq.~(\ref{eq:F}) becomes equivalent to (\ref{eq:Fcvm}), when the beliefs are equal to the marginal probability distributions. 

A complete parameterization of the beliefs in terms of connected correlation parameters $C$ is
\begin{equation}
  b_R(\sigma_R) = \frac{1}{2^{|R|}} \bigg[1 + \sum_{p \in P(R)} \prod_{s \in p} C_s \prod_{i \in s}\sigma_i  \bigg]\;, \label{eq:basCC}
\end{equation}	
where $P(R)$ are non-empty partitions over $R$, and $s$ are the elements (subsets) in each partition. Explicit examples for small subsets are given in Appendix \ref{app:beliefs2cc}, in the CVM framework any correlation parameter $C_s$ is zero unless $s \in R$ for some region $R$ in the approximation. The beliefs are by this choice normalized, and share parameters so as to be consistent on all marginals. We can interpret $b_R$ as locally consistent probabilities provided $0 \leq b_R(\sigma_R)\leq 1$.

The beliefs are, in a standard implementation, fixed by minimizing the free energy subject to local consistency requirements (maximum entropy). The beliefs determined in this way equal the marginal probabilities only for the special case of a junction tree. There are two possible sources of error in the approximate case: (A) region selection, (B) marginal distributions.

A junction tree is composed of large regions in many cases, making calculation of (\ref{eq:Scvm}) impractical even for known marginals. A compromize is to select only a subset of small regions. For graphs of special topology, or where out-of-region correlations are weak, these approximations are good, or occasionally asymptotically exact, and $b_R$ are close to the true marginal probabilities.

A hierarchy of mean-field approximations are recovered from CVM: The naive mean field (NMF) approximation is achieved by selecting single-variable regions, since $(i,j)$ are not contained in any regions $C_{ij}=0$ and $b_{ij}=b_i b_j$ for purposes of evaluating the energy (\ref{eq:Ecvm}), and (\ref{eq:Scvm}) simplifies to
\begin{equation}
  S_{N}(C) = - \sum_i \Tr \left[b_i \log b_i\right] \;. \label{eq:STAP}
\end{equation}
Note that in this expression, and henceforth, we write the variational dependence as $C$ rather than $b$, all beliefs are functions of $C$ through (\ref{eq:basCC}). 
The Bethe approximation includes NMF regions, and adds one pair region for each non-zero coupling ($J_{ij}\neq 0$). The entropy approximation $S_{B} = S_{N} + \Delta S_{B}$, where the correction to the entropy is
\begin{equation}
  \Delta S_{B}(C) = - \sum_{ij: J_{ij}\neq 0} \Tr \left[ b_{ij}\log\left(\frac{b_{ij}}{b_ib_j} \right)\right]\;.
  \label{eq:SBethe}
\end{equation}
Assuming small pair correlations we could consider an expansion of (\ref{eq:SBethe}) to quadratic order in the pair correlation parameters, from such an approach one can derive the TAP equations.
One possibility beyond Bethe (that we call the $P_X$ approximation) includes all plaquette regions up to maximum size $X$. A plaquette $P$ is a closed loop of coupled variables, without chords; an ordered set of $|P|$ variables ($i_1,\ldots,i_{|P|}$) such that $J_{i_{x-1},i_{x}}\neq 0$ (allowing $i_0=i_{|P|}$). The entropy is $S_{P_X} = S_{B} + \Delta S_{P_X}$, with
\begin{equation}
  \Delta S_{P_X}(C) = - \sum_{P} \Tr \left[b_{P}\log\left(\frac{b_{P}\prod_{x=1}^{|P|} b_{i_x}}{\prod_{x=1}^{|P|} b_{i_{x-1},i_{x}}} \right)\right]\;,
  \label{eq:SPlaquette}
\end{equation}
assuming plaquettes overlap on at most one edge.

A choice is made to approximate the entropy by a particular region selection, and the parameters (constrained beliefs, or equivalently correlation parameters) are fixed by minimizing the free energy, by 
\begin{equation}
  \left. \frac{\partial F_{CVM}}{\partial C_s}\right|_{C=C^*} = 0 \label{eq:mincond}\;,
\end{equation}
where the Hessian should also be positive definite. Linear response (LR) about this minimum then approximates the connected correlation on subset $s$ as
\begin{equation}
  \chi_s = \left.\frac{\partial^{|s|} F_{CVM}(C)}{\prod_{i \in s}\partial H_i}\right|_{C=C^*} \label{eq:linearresponse}\;.
\end{equation}
Except for cases where the free energy is exact the parameters $C_{s}^*$ and LR estimates $\chi_{s}$ disagree, on subsets $s$ of size $|s|>1$.
By a simple argument~\cite{Parisi:SFT} it is expected that parameters fixed by the saddle-point criteria (\ref{eq:mincond}) will be poorer estimates to the true correlations than those estimated by LR about the saddle-point (\ref{eq:linearresponse}). CVM is fundamentally a variational method, inducing an approximation to the probability measure $P(\sigma)=P_{exact}(\sigma)+\epsilon\,\delta P(\sigma)$. Since $O(\epsilon^d)$ errors arise in quantity determined by $d^{th}$ derivatives of the free energy, the correlation parameters determined by first derivative conditions (\ref{eq:mincond}) are of lower fidelity than the LR estimates obtained by higher order derivatives (\ref{eq:linearresponse}). However, by interpreting $\chi$ as the true correlations we implicitly accept the inaccuracy of $b_R$ as correct marginal probabilities at the saddle-point.

In our new method we require consistency between the two correlation estimates over a set $\Omega$
\begin{equation}
  C_{s}^* = \chi_{s}\;: \qquad s \in \Omega\;,\label{eq:linresponseidentity}
\end{equation}
The elements of $\Omega$ are subsets in atleast two indices, and these subsets must all be contained in some region forming the CVM approximation. 

If the approximation is NMF, consisting only of vertex regions, the set $\Omega$ must be empty and so we do not change the standard approximation. For purposes of this article we consider two simple modifications for the Bethe and Plaquette approximations: for the direct problem the set $\Omega$ is all edge regions with non-zero $J_{ij}$ ($\Omega=\{(i,j): J_{ij}\neq 0\}$); for the inverse problem the set $\Omega$ is of maximum size (all possible subsets of regions in the approximation without repetitions, $\Omega=\cup_R\{s: s \subset R, |s|\geq 2\}$). For the Bethe approximation both definitions of $\Omega$ are equivalent, for the Plaquette approximation we do not constrain three point (and higher order) connected correlations in the direct problem, whereas we do for the inverse problem. For the inverse problem we certainly want $\chi_s$ to equal $C_s$ for all $s$, we want to use the data to fix both correlations to the same value. For the direct problem implementation of the plaquette method we do not include the constraint on 3 point correlations for several reasons: brevity in explanation, technical convenience (it proves much simpler to fix 3-point correlations by maximum entropy than by the linear response identity), and the intuition that three point correlations will be less significant than two point correlations.

To implement correlation constraints we invoke a modified entropy approximation (\ref{eq:Scvm}), with slack parameters $\lambda$
\begin{equation}
  S_{\lambda}(C) = S(C) - \sum_{s \in \Omega}\lambda_s C_{s}\;. \label{eq:lamC}
\end{equation}
Choosing $\lambda$ to satisfy (\ref{eq:mincond}) leaves $C_{\Omega}=\{C_{s}:s\in \Omega\}$ unconstrained by the saddle-point equations, such that we can fix it self-consistently from (\ref{eq:linearresponse}) and (\ref{eq:linresponseidentity}). $\lambda$ may be large for poor region selection, and $\lambda=0$ when CVM is exact.

In Appendix \ref{app:Free_energy_modified_fields} we reformulate the free energy as a CVM approximation with modifications of the fields and couplings and additional variational parameters, this provides some additional insight into the complexity relative to a standard CVM implementation of the modification (\ref{eq:lamC}).

\section{A general framework for linear response}
In our method a minimum of the free energy is first determined.
The saddle-point equations, $\partial F_{CVM}/\partial C_i$ are
\begin{multline}
  0 = - \beta H_i - \sum_j \beta J_{ij}C_j + \atanh(C_i) + L_i(C)\;,\\
  L_i(C) = \sum_{R (\neq i): i \in R} c_R \Tr \left[
  \frac{\sigma_i}{2} b_{R\setminus i} \log \left( \frac{b_{R}}{b_i}\right)\right]\;,
  \label{eq:saddleM}
\end{multline}
where $b_{R\setminus i}$ is the belief over the region $R$ excluding $i$.
Eq. (\ref{eq:saddleM}) includes the NMF result and the correction $L_i(C)$.
The saddle-point equations, $\partial F_{CVM}/\partial C_s$ for larger sets $|s|>1$, are
\begin{equation}
  \beta J_{s} - \lambda_{s} = \!\!\!\!\sum_{R:s\in R}\!\! c_{R} \Tr\left[\prod_{i\in s}\left(\frac{\sigma_i}{2}\right) b_{R\setminus s}\log \left( b_{R}\right)\right]\;,
\label{eq:saddleC}
\end{equation}
with $J_{s}=0$ for sets $|s|>2$ and $\lambda_s= 0$ if $s \notin \Omega$. Eqs.~(\ref{eq:saddleM},\ref{eq:saddleC}) fix a subset of parameters $C,\lambda$: in our method we fix $\lambda$ and $C\setminus C_{\Omega}$, whereas the standard approach fixes all of $C$ (given $\lambda_s=0,\forall s$). More details of the derivation of the saddle-point equations (\ref{eq:saddleM}) and (\ref{eq:saddleC}) are  in Appendix \ref{app:saddles}.

Going forward we assume this is a well-defined minimum in the variational parameters ($C$): differentiable and not on the boundary of the feasible parameter space. In response to a variation in the fields $H_z\rightarrow H_z+ \delta H_z$ the parameters are perturbed $C_s\rightarrow C^*_s+\delta C_s$, and a quadratic order free energy describes the fluctuation
\begin{equation}
F_{CVM}(C^*) + \sum_i (C^*_i + \delta C_i)\delta H_i + \sum_{s,s'} \delta C_s Q_{s,s'} \delta C_{s'}/2 \label{eq:quadform}
\end{equation}
In this expansion we treat $\partial\lambda/\partial H$ as zero, i.e. $\lambda$ is not a variational parameter.
The saddle-point equations are then 
\begin{equation}
  \sum_s Q_{i,s} \delta C_s =\delta H_i\;;\qquad \forall i\;,
\end{equation}
and 
\begin{equation}
  \sum_s Q_{s',s} \delta C_s = 0\;;\qquad \forall s' : |s'|>1\;.
\end{equation}
Solving this system of linear equations in $\delta C_{i}$, and identifying $\delta C_i = \beta \sum_j \chi_{ij}\delta H_j$ as linear responses, we can write a system of $N^2$ equations
\begin{equation}
[\chi^{-1}]_{i,j} = - \beta J_{ij} + \Phi_{i,j}(C^*)\;.
\label{eq:invC}
\end{equation}
The detailed relationships between the Hessian $Q$, $\Phi$ and some other linear response identities are described in Appendix \ref{app:HessianAndPhi}. 

For our choice of $\Omega$ in the direct problem Eqs.(\ref{eq:linresponseidentity},\ref{eq:saddleM},\ref{eq:saddleC},\ref{eq:invC}) form a closed set for the determination of $C$ and $\lambda$ given $J$ and $H$. Requiring consistency with respect to different choices of $\Omega$, e.g.\ 3-point correlations, is possible, but requires higher order derivates. 

For the inverse problem we do constrain all higher order (more than 2-point) correlations in the case of the Plaquette approximation, but note that for any CVM approximation Eqs.(\ref{eq:linresponseidentity},\ref{eq:saddleM}) and the off-diagonal component of (\ref{eq:invC}) already form a closed set of equations for the determination of $H$ and $J$ given $C$. We need not know either $\lambda$, or the structure of the equation determining the higher order responses, in order to complete the inference of $J$ and $H$. 

\section{Comparison of methods}

Eq.~(\ref{eq:invC}) applies for any $\lambda$ value. Standard ($\lambda=0$) LR results are reproduced by solving (\ref{eq:saddleM},\ref{eq:saddleC},\ref{eq:invC}) without requiring (\ref{eq:linresponseidentity})~\cite{RicciTersenghi:InverseIsing,Kappen:BML}.

In our new approach, beginning from the Bethe approximation, we determine the on-diagonal elements in (\ref{eq:invC}) to be
\begin{equation}
  \Phi^{B}_{ii} = \frac{1}{1-C_i^2} \left[1 + \sum_{j: ij \in \Omega} \frac{C_{ij}^2}{(1-C_i^2)(1-C_j^2)-C_{ij}^2}\right]\;.\label{eq:BeliefOnDiagonal}
\end{equation}
The entropic off-diagonal matrix components, for edge regions $\{ij : J_{ij}\neq 0\}$, are
\begin{equation}
\Phi^{B}_{ij} = J^{IP}(C_{ij},C_i,C_j) - \frac{C_{ij}}{(1-C_i^2)(1-C_j^2)-C_{ij}^2}\;,\label{eq:BeliefOffDiagonal}
\end{equation}
where other off-diagonal components are zero. $J^{IP}$ is the independent pair approximation
\begin{equation}
  J^{IP}(C_{ij},C_i,C_j) = \Tr \left[\frac{\sigma_i \sigma_j}{4} \log b_{ij}\right]\;.
\end{equation}
More details are provided in Appendix \ref{app:SessakMonasson}.
Using $\Phi_{ij}^{B}$ in (\ref{eq:invC}) the Sessak-Monasson~\cite{Sessak:SME} result for small correlation parameters is recovered. For the plaquette method we have corrections $\Phi^{P}=\Phi^B + \Delta \Phi^{P}$. In the simplest case with triangular plaquettes, and spin-symmetric probabilities ($C_i=0$,$C_{ijk}=0$), we have a correction to the non-zero off-diagonal elements given by
\begin{multline}
 \!\!\!\Delta \Phi^{P_3}_{ij} \!=\!\! \sum_{k (\neq i,j)} c_{ijk}\left\lbrace  \frac{1}{4} \sum_{a=\pm 1} \!\!a \log\left(1 \!-\! \frac{(C_{jk} + a C_{ik})^2}{(1+ aC_{ij})^2}\right) \right. + \\
\!\!\!\left.\frac{(C_{ik} - C_{jk} C_{ij}) ( C_{jk} - C_{ik} C_{ij})}{(1 - C_{ij}^2) (1 - C_{jk}^2 - C_{ik}^2 + 2 C_{jk} C_{ik} C_{ij} - C_{ij}^2)}
 \right\rbrace\;,
\label{eq:DeltaPhiP3}
\end{multline}
where $c_{ijk}=1(0)$ for included (excluded) plaquettes, as described in Appendix \ref{app:PhiP3}.

\subsection{High temperature expansions}
For a fully connected model we can consider the leading order errors in the high temperature regime by an expansion in $\beta$, as described in Appendix \ref{sec:hightempexpansion}. For the symmetric case, an approximation inclusive of all edge regions yields a correction in $\Phi_{ij}$
\begin{equation}
  \Phi^B_{ij} - \Phi_{ij}^{exact} = - \beta^5\sum_{k (\neq i,j)}2 J_{ij} J_{ik}^2J_{jk}^2 + O(\beta^6)\;,
  \label{eq:beta5}
\end{equation}
to be compared to the $O(\beta^4)$ error in standard Bethe. For the Plaquette approximation, including all triplet regions, the error is improved to $O(\beta^6)$. In the special case of zero external field (a symmetric solution) it is
\begin{multline}
\Phi^{P_3}_{ij} - \Phi_{ij}^{exact} = \beta^7\!\!\!\!\!\!\sum_{k<l (\neq i,j)}\!\!\!\!\!\! 2 J_{kl} \big[J_{ij} J_{kl} (J_{ik} J_{jl} + J_{il} J_{jk})^2 +\\
2 J_{ik} J_{jk} J_{il} J_{jl}
(J_{ij} J_{kl} + J_{ik} J_{jl} + J_{il} J_{jk}) \big] + O(\beta^8)\;,
\label{eq:beta7}
\end{multline}
to be compared to $O(\beta^5)$ errors for the standard LR implementation.
The errors in (\ref{eq:beta5}) and (\ref{eq:beta7}) are evaluated given the exact value of $C$, which is pertinent to an idealized inverse problem application. If instead we consider the direct problem we are more interested in errors on the statistics $C$; these errors depends on the on-diagonal component error of $\Phi$, which is unimproved in the new method. In~\cite{Raymond:Correcting} an analysis and remedy is proposed that involves including {\em on-diagonal} constraints, as discussed in Appendix \ref{app:ondiagoffdiag}. 

\subsection{Direct problem}

The direct problem of determining magnetizations $\{C_i\}$ and correlations (e.g. $\{C_{ij}\}$) given $H$,$J$ requires the simultaneous solution to (\ref{eq:linresponseidentity},\ref{eq:saddleM},\ref{eq:saddleC},\ref{eq:invC}). A possible iterative scheme for NMF, Bethe and Plaquette approximations is
\begin{eqnarray}
  C_i^{t+1} &\leftarrow& \tanh\biggr[\beta\biggr(H_i + \sum_j J_{ij} C_j^t\biggr) - L_i(C^t)\biggr]\;,\label{eq:linrespiter1}\\\
  C_{ij}^{t+1} &\leftarrow&  \chi_{ij}^{t} =\big[\left(-\beta J + \Phi(C^t)\right)^{-1}\big]_{ij}\;, \label{eq:linrespiter2}\\
  b_P^{t} &\leftarrow& \!\!\mathrm{argmin}\biggr\{\!\Tr \left[b_P^t \log b_{P}^t \bigr|\; \{C_i^t\},C_\Omega^t\right]\biggr\} \label{eq:linrespiter3}\;.
\end{eqnarray}
All approximations (NMF, Bethe, $P_X$) require (\ref{eq:linrespiter1}), only Bethe and $P_X$ approximations ($\lambda\neq 0$) require (\ref{eq:linrespiter2}), and only the $P_X$ approximations require (\ref{eq:linrespiter3}) where $b_P^t$ is the belief parameterized by the correlations $\{C_s^t: s \in P\}$. 
Thus (\ref{eq:linrespiter3}) assigns the maximum entropy estimate to all connected correlations on a plaquette region $P$ not fixed by (\ref{eq:linrespiter1}) and (\ref{eq:linrespiter2}). Eq. (\ref{eq:linrespiter3}) is an easily solved local convex optimization, subject to linear constraints determined by $C^t_{\Omega}$ and $\{C^t_{i}\}$.
At sufficiently high temperature the scheme is convergent, and the solution stable. However, at lower temperatures the process may be unconvergent, and so strong damping and/or special update ordering is required. We describe in more detail solving the equations for the special case of homogeneous solutions on a lattice in Appendix \ref{app:HoLatt}.

\begin{figure}[!ht]
  \includegraphics[width=\columnwidth]{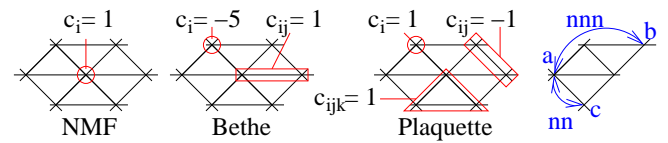}
  \caption{\label{fig:trilat} (color online) Regions and counting numbers for a triangular lattice. $\{a,c\}$ are nearest neighbors (nn), $\{a,b\}$ are next-nearest neighbors (nnn). We abbreviate $\{\chi_{nn},C_{nn},\lambda_{nn}\}$ and $\chi_{nnn}$ for the corresponding homogeneous quantities.}
  \vspace{-0.15in}
\end{figure}

We find our method to be promising for models with many short loops. The homogeneous triangular lattice model (HTL) with $H_i=0$, $J_{ij}=1$ for nearest neighbors and $0$ otherwise, is a well understood canonical model that has a ferromagnetic transition point at $\beta_c=0.275$ \cite{Baxter:ESM}, and for $\beta<0$ is fully frustrated with no long range order, but a Kosterlitz-Thouless transition~\cite{Wannier:A,Stephenson:IMSC}. For finite lattice implementations we choose periodic boundary conditions (periodicity $L$). Figure \ref{fig:trilat} shows the corresponding region based approximations in the direct problem.

\begin{figure}[t]
\setlength{\unitlength}{1.4mm}
\begin{picture}(60,45)
\put(0,0){\epsfysize=45\unitlength\epsfbox{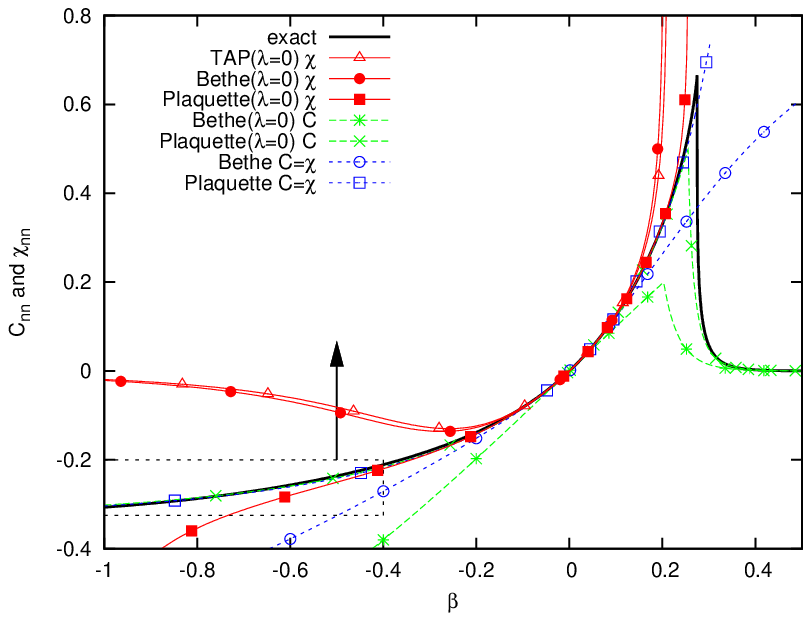}}
\put(6,18){\epsfysize=23\unitlength\epsfbox{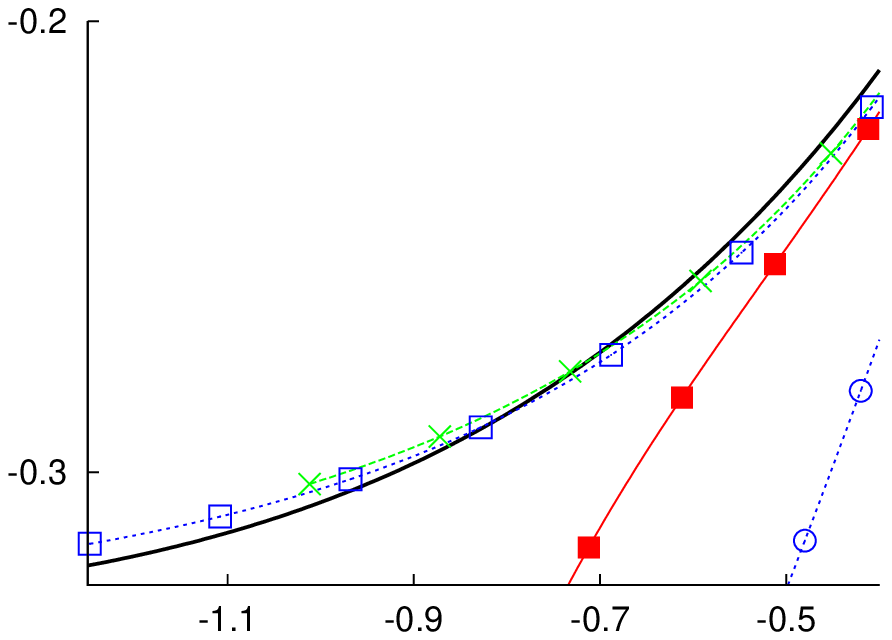}}
\end{picture}
\caption{\label{fig:TrianglularLattice} (color online) Nearest neighbors correlation estimates for the asymptotic ($L\rightarrow \infty$) HTL. Magnetized branches are shown only for Bethe and Plaquette ($\lambda=0,\beta>0$) correlation parameters.}
\end{figure}

\begin{figure}[tbh]
\setlength{\unitlength}{1mm}
\includegraphics[width=0.49\columnwidth]{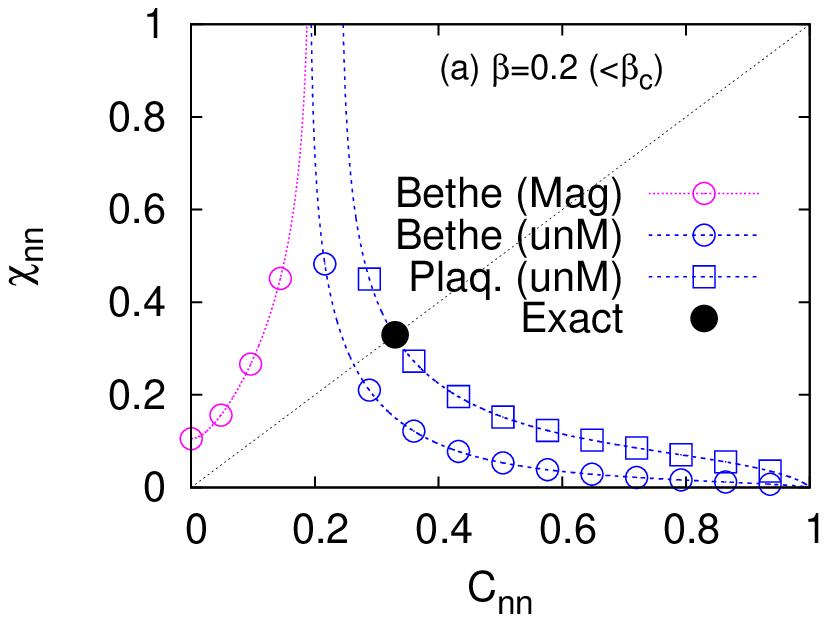}
\includegraphics[width=0.49\columnwidth]{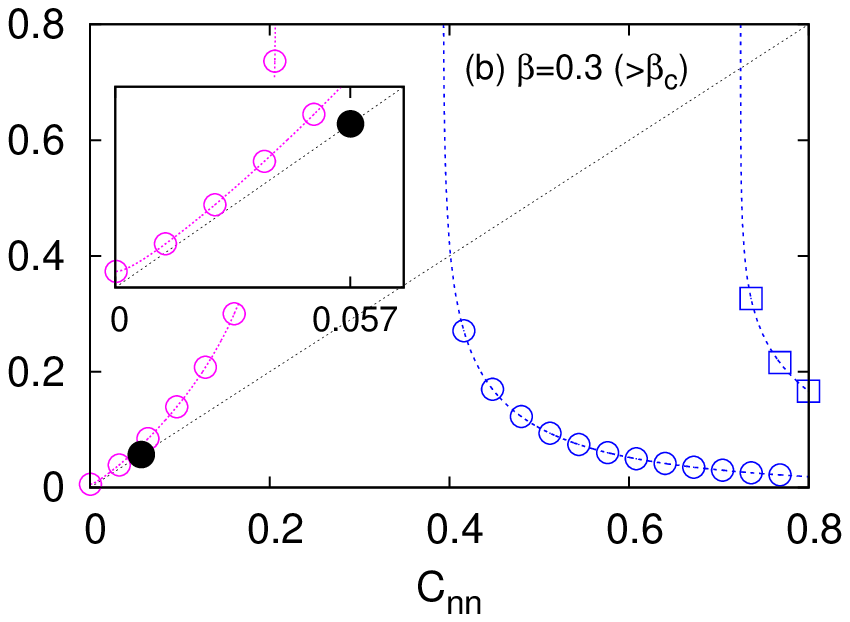}
\caption{\label{fig:FixedPointBehaviour} (color online) The response as a function of the correlation parameter for the asymptotic ($L\rightarrow \infty$) HTL with (left) $\beta<\beta_c$,(right) $\beta>\beta_c$. Inset (right) shows a magnified version of the magnetized branch. Bethe (magnetized and unmagnetized) and $P_3$ (unmagnetized) approximation results are shown. Our method requires the intersection point ($\chi_{nn}=C_{nn}$), to be compared with the exact solution.}
\end{figure}

Fig.~\ref{fig:TrianglularLattice} shows nearest neighbor correlation estimates obtained in the thermodynamic limit by Fourier techniques, as described in Appendix \ref{app:HoLattasy}. We show the exact result by the black line, standard LR methods in red (label $\chi$), standard methods minimizing F in the variational parameters in green (label $C$) and our new method in blue. All methods perform well at high temperature (small $|\beta|$), and magnetized solutions are accurate for $\beta \gg \beta_c$. Standard methods undergo spurious continuous transitions for $\beta\lesssim \beta_c$, and the NMF and Plaquette approximations also undergo a transition in the frustrated regime $\beta<0$. LR estimates diverge at these spurious critical points. The standard ($\lambda=0$) $P_3$ method performs well in the estimate of $C_{nn}$ (nearest neighbor correlation), for the unmagnetized solutions, but only in the stable range $\beta \in (-1.01,0.255)$, while our new $P_3$ method performs well in the entire frustrated region and up to the true critical temperature $\beta<\beta_c$ (see the inset of Fig.~\ref{fig:TrianglularLattice}). However, the unmagnetized solution does not exhibit continuous phase transitions for the new methods for $\beta \sim \beta_c$, as it should. At low temperature convergence problems hinder the construction of solutions (iteration of \ref{eq:linrespiter2} fails due to large gradients, as shown figure \ref{fig:FixedPointBehaviour}), the unmagnetized $P_3$ solution is constructed only for $\beta<0.3$. Certainly the $P_3$ unmagnetized solution is unfeasible for $\beta>0.35$ (already significantly below the critical temperature), since the Hessian becomes singular for any $C_{nn}<1$, the unmagnetized Bethe solution is stable to much larger $\beta$ as shown.

\begin{figure}[t]
\begin{center}
\includegraphics[width=0.49\columnwidth]{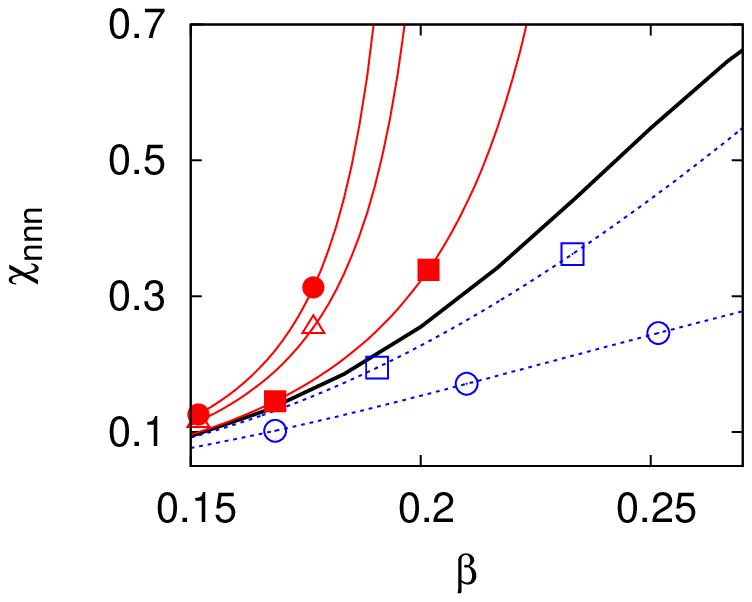}
\includegraphics[width=0.49\columnwidth]{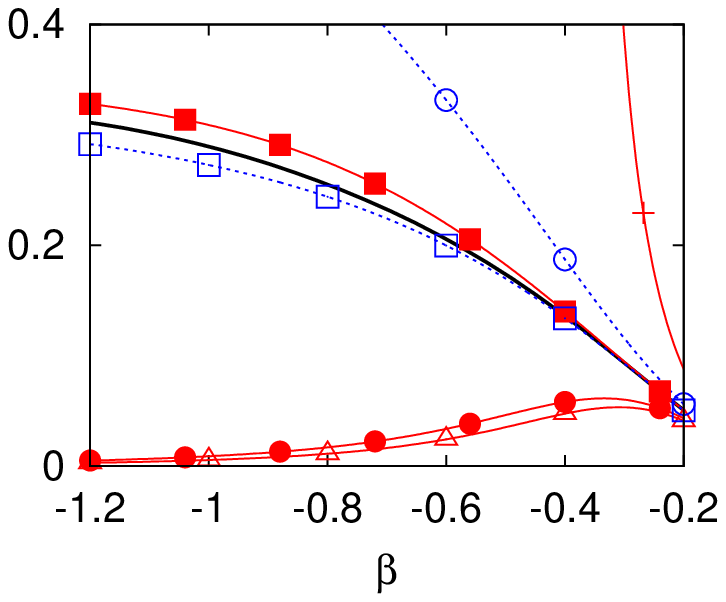}
\end{center}
\vspace{-0.25in}
\caption{\label{fig:Cnnn} (color online) Next nearest neighbor correlations for the HTL $L=5$, unmagnetized branches. Curves as figure \ref{fig:TrianglularLattice}.}
\end{figure}

Lattice models with finite $L$ do not strictly speaking exhibit a phase transition, but many phenomena are well described by models with this feature.
However, data collected in real applications often do not show any phase transition phenomena~\cite{Morcosa:DCA,Cocco:NC}; a more general test of inference methods is the quality of the marginals predicted.

Correlations that extend beyond the approximation regions are not amongst the CVM parameters, LR is required to determine pair correlations at distance larger than $1$. For $\beta\in (-\infty,\beta_c]$ the new method improves upon standard implementations for many significant terms in $\chi$ and $\chi^{-1}$. Figure \ref{fig:Cnnn} shows the next nearest correlations calculated on a finite model $L=5$: the new method estimates are superior to their counterparts for most $\beta$. The values calculated for $L=5$ are close to those for $L\rightarrow \infty$ for $\beta<\beta_c$, although in the case of $L=5$ the tripartite lattice symmetry is broken so that for $\beta<0$ there is extra frustration, and the unmagnetized solutions are more stable. The NMF unmagnetized solution is unstable for $\beta<-0.382$ 
, 
but other unmagnetized solutions are stable for $-1.2<\beta$. An interesting feature of the new method is that it overcompensates the error of the standard method; so a combination of the two can lead to even better results. 

\begin{figure}[!ht]
  \includegraphics[width=\columnwidth]{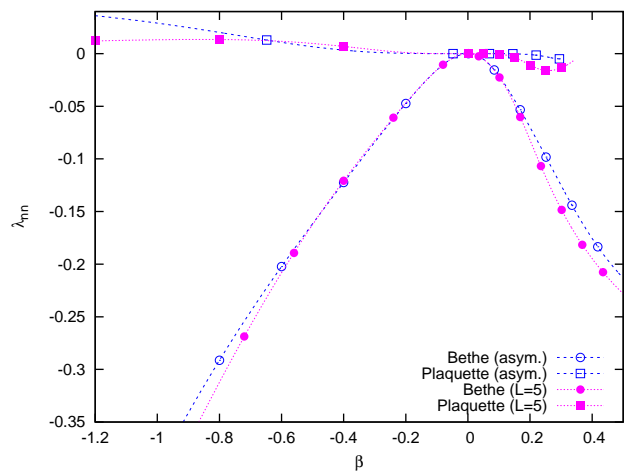}
  \caption{\label{eq:lambdavals} (color online) Values of $\lambda_{nn}$ that achieve the curves of figure \ref{fig:TrianglularLattice} and \ref{fig:Cnnn} in the Bethe and Plaquette approximations with $L=5$ and $L\rightarrow \infty$ (asymptotic).}
  \vspace{-0.15in}
\end{figure}
We can see that the LR ($\chi_{nn}$) and max-entropy ($C_{nn}$) estimates are for the Plaquette approximations much closer, and relatively accurate, compared to those for Bethe and NMF. Correspondingly the values of $\lambda_{nn}$ are much smaller in absolute value in the Plaquette approximation as shown in figure \ref{eq:lambdavals}. 
For the asymptotic curves ($L\rightarrow \infty$) we observe a monotonic trend in $\lambda_{nn}$, for the finite model ($L=5$) the effect of the boundary causes the curve to become non-monotonic. However, note that for $L=5$ the effective potential ($\beta-\lambda_{nn}$, discussed Appendix \ref{app:Free_energy_modified_fields}) remains monotonic as a function of $\beta$ for all solutions presented. In the ferromagnetic region ($\beta>0$) the value of $\lambda_{nn}$ is negative for both approximations, which has the effect to reduce the ferromagnetic susceptibility of the unmagnetized solution.

In all the results discussed so far we have used an annealing technique to obtain the curves, so the curves presented are those that are obtained continuously from the unique high temperature solution to the equations. These are all unmagnetized solutions. In the zero field lattice models tested we found that for large $\beta$ either a magnetized solution appeared discontinuously, or that no magnetized solution exists.

As shown in figure \ref{fig:FixedPointBehaviour} we plot the curves for the linear response correlation estimate $\chi_{nn}$ against the parameter value in the asymptotic ($L\rightarrow \infty$) HTL; a similar pattern of curves applies for a variety of translationally invariant lattice models we solved. We show the magnetized branch only for the Bethe approximation, since the $P_3$ approximation cannot be solved asymptotically in the magnetized case by the Fourier method outlined in Appendix \ref{app:HoLattasy}. Our method dictates solutions according to (\ref{eq:linresponseidentity}), i.e. the point(s) $\chi_{nn} = C_{nn}$, which can be achieved by a choice of $\lambda_{nn}$. 

For $\beta<\beta_c$ the true solution for a lattice model is unmagnetized. Our approximations also determine unique unmagnetized solutions that slightly underestimate the true value, a typical case is figure \ref{fig:FixedPointBehaviour}(a). For frustrated lattice models, such as the HTL with $\beta<0$, the unmagnetized solution is also found to be unique, and the curve for $-\chi_{nn}$ versus $-C_{nn}$ is similar to the unmagnetized branch of figure \ref{fig:FixedPointBehaviour} (with the magnetized branch absent).

For $\beta>\beta_c$ the true solution is magnetized and the connected correlation decreases from a peak at $\beta_c$. Our method typically exhibits an unmagnetized solution for $\beta \gtrsim \beta_c$, with large connected correlation, a typical case is \ref{fig:FixedPointBehaviour}(b). This solution can persist to very large $\beta\gg\beta_c$ as shown in figure \ref{fig:TrianglularLattice}. Alongside this we typically find either a magnetized solution, or magnetized pseudo-solution ($\chi_{nn}\approx C_{nn}$), for which the marginals are better estimated. A typical pseudo-solution behaviour is shown figure \ref{fig:FixedPointBehaviour}(b,inset). The pseudo-solution behaviour we found to be typical of standard 2D lattice models. By contast moving to the 3D model we found for $\beta\gtrsim \beta_c$ the coexistence of stable magnetized and unmagnetized solutions -- the figure is similar to \ref{fig:FixedPointBehaviour}(b) except that the magnetized curve crosses at two points, to give one locally stable magnetized solution alongside the locally stable unmagnetized solution. A discontinuous transition from the paramagnetic to the ferromagnetic solution is apparent in these cases. 

The curves of figure \ref{fig:FixedPointBehaviour} also dictate the dynamics of the iterative procedure (\ref{eq:linrespiter2})) with other terms (\ref{eq:linrespiter1},\ref{eq:linrespiter3}) at their fixed point values. Due to the large gradient at the unmagnetized fixed point strong damping is required in the proposed iterative method to find the unmagnetized solutions for large $\beta$ (large $|\beta|$ in the case of frustrated regimes). By contrast, the magnetized solution (or pseudo-solution, if we allow $\chi_{nn}\sim C_{nn}$) has benign dynamics.

\subsection{Inverse problem}

\begin{figure}[t]
  \begin{center}
    \includegraphics[width=\columnwidth]{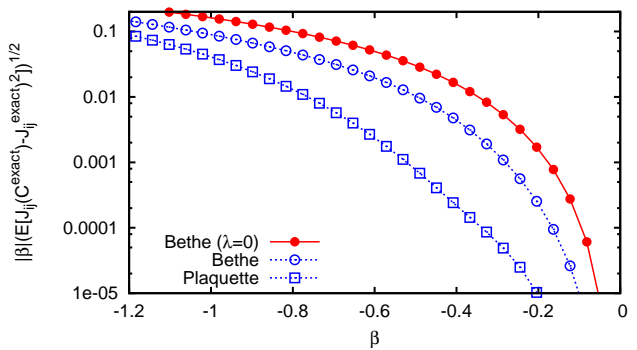}
  \end{center}
  \vspace{-0.25in}
  \caption{(color online) THL: error in inference of $J$ from exact statistics.}
  \label{fig:triangle25}
\end{figure}

\begin{figure}[t]
  \begin{center}
    \includegraphics[width=\columnwidth]{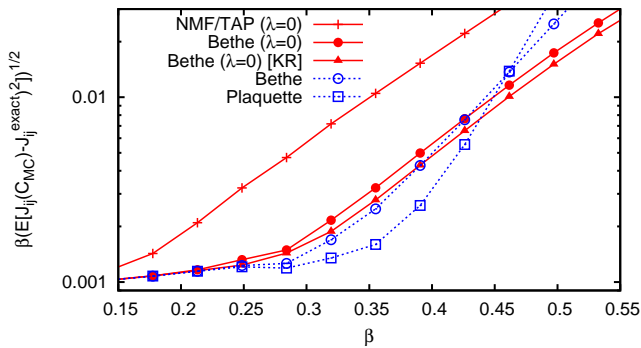}
  \end{center}
  \vspace{-0.25in}
  \caption{(color online) Error in inferring couplings $J$ for a diluted 2D square ferromagnet, from statistics of $10^6$ independent samples. [KR] employs the Kappen-Rodriguez normalization \cite{Kappen:BML}.}
  \label{fig:dilutedferro49}
\end{figure}

A simpler application of our method is for the inverse problem: given sample statistics, determine $J$ and $H$~\cite{Aurell:IIP,RicciTersenghi:InverseIsing}. With ignorance of the distribution of couplings (and topology), we must have unbiased region selection: all edges for Bethe, and all (triangular) Plaquettes for $P_3$.  In the new method we take $C$ and $\chi$ equal to the correlation statistics and solve first (\ref{eq:invC}) in the off-diagonal elements for $J_{ij}$, and then (\ref{eq:saddleM}) for $H$.  Eq. (\ref{eq:saddleC}) can be used to determine $\lambda$, which would be a measure of model fidelity. In standard mean field methods the same assumptions are made on region selection, but only $\chi$ and $\{C_i\}$ are determined from the statistics, all other $C$ obey the saddle point equations (\ref{eq:saddleC}) with $\lambda=0$ (thus making equations solvable for Bethe and TAP~\cite{RicciTersenghi:InverseIsing}).

Figure \ref{fig:triangle25} demonstrates the results for estimation of matrix $J$ in the HTL $L=5$ based on exact data. The improved scaling at small $|\beta|$ is as anticipated in equations~(\ref{eq:beta5}), (\ref{eq:beta7}). However, even at low temperature reconstruction is significantly improved by the new methods. Although $\Phi_{nn}$ determines the error, note that the approximation is different to that used in the direct problem: the 2D triangular structure is discovered, unlike in the direct problem where it is assumed in the region selection.

Figure \ref{fig:dilutedferro49} demonstrates results for an instance of a $7$ by $7$ diluted square lattice Ising model in zero field. Each coupling is assigned according to the probability distribution $P(J) = 0.7\delta_{J,1} + 0.3 \delta_{J,0}$. The reconstruction assumes $H_i=0$, but no knowledge of $J$. We generated the pair-correlation matrix from independent Monte Carlo measures.  Sampling errors limit all methods for small $\beta$. When $\beta$ is large enough the error of the method exceeds the sampling one. A $\beta$ interval exists in which the new methods improve over standard ones. The triangular Plaquette approximation improves over Bethe, despite the absense of triangles in the model (the shortest loop is of length 4). For larger $\beta$ the model undergoes a rapid growth in correlation length, far beyond the edge/triangular regions selected, all mean-field methods are prone to significant errors.

Since at the Bethe level our method coincides with the Sessak-Monasson expression it is not a surprise that we outperform other mean field methods at high temperature due to the improved scaling (\ref{eq:beta5}). The plaquette approximation is by no means guaranteed to outperform the Sessak-Monasson expression outside the weak coupling regime where a superior (\ref{eq:beta7}) scaling applies, but this advantage persists at intermediate temperatures for the two models presented. Realization of the high temperature scaling is only feasible if data is of very high quality, in practical applications this is unlikely, sampling will be subject to errors and performance may not be significantly improved over NMF, the more interesting comparison of methods is for intermediate and low temperatures. 

We studied for example the 2D square lattice Edward Anderson model with zero field and $J=\pm 1$ coupling distribution across a range of system sizes $L=4$ to $L=32$ at intermediate temperatures $\beta \in [0.3,0.8]$ and a small number of samples for each system size~\cite{Edwards:TS,Haijun:EAData}. We found that the $\lambda\neq 0$ outperform standard implementations of Bethe and NMF in all samples, however for nearest couplings we found that the Bethe ($\lambda\neq 0$) method provided a marginally superior estimate (i.e. estimates of $J_{ij}$ were closer to $1$ in absolute value for nearest neighbors $i,j$), whilst the $P_3 (\lambda\neq 0)$ approximation provided a marginally superior estimate for the absent couplings (i.e. estimates were closer to $0$ for $J_{ij}$, where $ij$ are not nearest neighbors). We will present more detailed analysis of the inverse problem across a range of problems in a forthcoming paper.

\section{Discussion and Conclusion}

We propose a minimal modification to the mean-field free energy functional in order to make max-entropy estimates of correlations consistent with LR ones, in other words the Hessian consistent with the location of the free energy minimum. To do this we introduce a new set of parameters ($\lambda$) and constraints (\ref{eq:linresponseidentity}), and argue that this may move belief estimates closer to the true marginals, and hence the free energy estimate is improved. An alternative argument is that physical quantities, as measured by different combinations of derivatives of the variational free energy, should be consistent at a good evaluation point, as discussed in Appendix \ref{app:ondiagoffdiag}. 

For the direct problem the value of $\lambda$ is unique and can be found by expansion in the weak coupling regime, but even for simple models we find a solution meeting criteria (\ref{eq:linresponseidentity}) may not exist at low temperature; this is in contrast to standard implementations of CVM where minima always exist~\cite{yedidia}. The absence of a continuous transition to a magnetized solution in some ferromagnetic lattice models might be viewed as a pathology, although the standard mean field exponent $1/2$ is itself a significant underestimation of the abruptness of these transitions. We have proposed a strict implementation of the condition (\ref{eq:linresponseidentity}), but by relaxing this condition slightly we might find better solutions such as the magnetized pseudo-solution for the HTL. By contrast, for the inverse problem there always exists a unique set of values $J$,$H$ solving (\ref{eq:invC},\ref{eq:saddleM},\ref{eq:linresponseidentity}), which determines unique values $\lambda$ through the saddle-point equations (\ref{eq:saddleC}); this in an improvement relative to the standard implementation of CVM methods for the inverse problems where solutions may not exist for certain models~\cite{RicciTersenghi:InverseIsing}.

In our framework we are not able to produce a general argument that guarantees our new variational approach produces either an upper or lower bound to the free energy (standard CVM has the same problem), and more importantly we cannot provide guarantees of feasibility or uniqueness of $\lambda$. Despite this it seems that the new method is very effective in both the inverse and direct problems, indicating the constraints we are introducing are beneficial as extentions of variatonal frameworks; and we maintain the property of CVM that when the region selection is correct the exact free energy is found by minimization ($\lambda=0$ is a solution). The framework is inclusive of previous NMF and Bethe approaches for the inverse problem, whilst providing a new variational basis for the Sessak-Monasson expression.

Another important set of issues are algorithmic, when feasible values of $\lambda$ exist how can they be found, and how can the free energy even be minimized given fixed $\lambda$. The constraints we introduce to the CVM are linear in the connected correlation parameters but non-linear in the belief paramaterization as discussed in Appendix \ref{app:Free_energy_modified_fields}. This restricts the class of methods available for (local or global) minimization, we have presented a simple iterative algorithmic method which is sufficient for high temperature, but leaves room for improvement. We have been able to develop message passing equations for our framework, which will be presented alongside the analysis of the direct problem in a broader class of models in a forthcoming work. Another important topic not covered in this paper are on-diagonal constraints~\cite{Raymond:Correcting,Yasuda:SPDC}, discussed in Appendix \ref{app:ondiagoffdiag}. These are important for improving estimation in the direct problem, and the estimation of the magnetizations in the inverse problem and can be straightforwardly incorporated in our framework.
\\
\begin{acknowledgments}
This work is supported by the Italian Research Minister through the FIRB Project No. RBFR086NN1XYZ . We extend thanks Aurelien Decelle for useful discussions. We thank Haijun Zhou for providing data on the 2D EA model.

\end{acknowledgments}
\bibliography{Bibliography}

\appendix 

\section{Beliefs as functions of correlations}
\label{app:beliefs2cc}
For the derivations restricted to NMF, Bethe and $P_3$ approximations we indicate the explicit forms for the beliefs in terms of the connected correlation parameters (\ref{eq:basCC}).
\begin{eqnarray}
  b_{i} &=& \frac{1+C_i\sigma_i}{2} \label{eq:bi} \;; \\
  b_{(i_1,i_2)} &=& b_{i_1 i_2} = \prod_{x=1}^2 b_{i_x} + C_{i_1 i_2}\prod_{x=1}^2 \left[\frac{\sigma_{i_x}}{2}\right] \label{eq:bij} \;\\
  b_{(i_1,i_2,i_3)} &=& b_{i_1 i_2 i_3} = \prod_{x=1}^3 b_{i_x} + C_{i_1i_2i_3}\prod_{x=1}^3 \left[\frac{\sigma_{i_x}}{2}\right] \label{eq:bijk}\\ &+& \sum_{x=1}^3 C_{i_1i_2i_3\setminus i_x}(C_{i_x}+\sigma_{i_x})\prod_{y=1}^3 \left[\frac{\sigma_{i_y}}{2}\right]\nonumber \;.
\end{eqnarray}
Note that unlike the representation of the beliefs in terms of the full correlations, the representation in terms of connected correlations is non-linear. Linearity and convexity are not preserved properties in the new basis, in particular the entropy term $\Tr[b_R \log b_R]$ is not a convex function of $C$ in general. However, since the parameters are variational, and the correlation parameters span the same space of beliefs, the free energy estimate is unaltered.

In the CVM framework parameters are only non-zero if they are subsets of some region in the approximation. Hence for the NMF approximation (\ref{eq:bij}) and (\ref{eq:bijk}) reduce to products of the marginal probabilities (\ref{eq:bi}).

\section{Derivation of saddle-point equations: (\ref{eq:saddleM}),(\ref{eq:saddleC})}
\label{app:saddles}
The Hamiltonian of the main text (\ref{eq:Hamiltonian}) is a special case of  
\begin{equation}
  H = - \sum_s J_s \prod_{i \in s} \sigma_i \label{eq:HamiltonianGeneral}
\end{equation}
where in the main text $J_i=H_i$ (the external fields), and $J_s=0$ for sets larger than $2$ (couplings are pairwise).

The free energy components in a generic CVM approximation become
\begin{equation}
  E(J,H,b) = - \beta \sum_s J_s \Tr \left[b_s\prod_{i\in s}\sigma_i\right]\;,
\end{equation}
and
\begin{equation}
  S(b) = \sum_R c_R \Tr \left[b_R \log b_R\right] + \sum_{s\in \Omega} \lambda_s C_s\;.
\end{equation}
$R$ are the regions forming the CVM approximation, $s$ are subsets of variables, and $\Omega$ is the set for which we require $\chi_s=C_s$. 

The saddle-points and Hessian are determined by derivatives with respect to the variational parameter. A useful identity given our parameterization is
\begin{equation}
  \frac{\partial b_R}{\partial C_s} = \prod_{i\in s}\left[\frac{\sigma_i}{2}\right] b_{R\setminus s}\;,
\end{equation}
where $R\setminus s$ is the set complement. The belief over the empty set is defined to be $b_{s\setminus s}=1$ and we define $b_{R\setminus s}=0$ for cases where $s$ is not entirely inside $R$ ($s\setminus R$ is not empty).

The saddle-point equation for any correlation parameter $\partial \beta F_{CVM}/\partial C_{t}$ is thus 
\begin{multline}
  0 = - \beta \sum_{s:t \subset s} J_s \Tr \left[\prod_{i\in t} \left[\frac{\sigma_i}{2}\right] b_{s\setminus t}\prod_{i\in s}\sigma_{i}\right] + \label{eq:saddle}\\
  \sum_{R: t \subset R} c_R \Tr \left[\prod_{i\in t} \left[\frac{\sigma_i}{2}\right] b_{R\setminus t} \log b_R\right] + \sum_{s \in \Omega} \delta_{t,s}\lambda_s  \;.
\end{multline}
For $t$ in atleast two indices this is identical to (\ref{eq:saddleC}) 
 recognizing that the first term reduces to $-\beta J_{ij}$ for pairwise couplings. 
We arrive at (\ref{eq:saddleM}) 
from (\ref{eq:saddle}) employing the counting numbers identity $\sum_{R: i \in R}c_i =1$ true for any $i$, and $\Tr [\sigma_i b_{i \cup R} \log b_i] = \Tr [\sigma_i b_i\log b_i]$ for any set $R$. In this way 
\begin{multline}
  \sum_R c_R \Tr \left[\frac{\sigma_i}{2} b_{R\setminus i} \log b_R\right] = \Tr \left[\frac{\sigma_i}{2} \log b_i\right]  + \\
 \sum_{R\setminus i} c_R \Tr \left[\frac{\sigma_i}{2} b_{R\setminus i} \log \left(\frac{b_R}{b_i}\right)\right]\;,
\end{multline}
thereby separating the naive mean field term for $t=i$ from corrections in (\ref{eq:saddleM}).

\section{Derivation of Hessian and $\Phi$ (\ref{eq:quadform}) and (\ref{eq:invC}) and an alternative response identity.}
\label{app:HessianAndPhi}
Following the derivation of the saddle-point equations we construct the Hessian by a derivative of (\ref{eq:saddle}) with respect to $C_u$
\begin{multline}
 \!\!\!Q_{t,u}\! = -\! \beta \!\!\!\sum_{s:t,u \in s} J_{s} \Tr \left[\prod_{i\in t} \left[ \frac{\sigma_i}{2}\right]\prod_{i\in u} \left[ \frac{\sigma_i}{2}\right] b_{(s\setminus t)\setminus u} \prod_{i \in s}\sigma_{i}\right] + \\
 \sum_{R: t,u \in R} c_R \Tr \left[\prod_{i\in t} \left[ \frac{\sigma_i}{2}\right]\prod_{i\in u} \left[ \frac{\sigma_i}{2}\right]  b_{(R\setminus t)\setminus u} \log b_R \right]  + \\
 \sum_{R: t,u \in R} c_R \Tr \left[\prod_{i\in t} \left[ \frac{\sigma_i}{2}\right]\prod_{i\in u} \left[ \frac{\sigma_i}{2}\right] \frac{b_{R\setminus t}b_{R\setminus u}}{b_R} \right]\;.
\end{multline}
The Hessian should be positive definite, for a consistent method, this can be checked.
We separate the matrix into blocks (in the case of NMF we have only the block $Q^{(1)}$ and we can take $\Phi=Q$ in that special case),
\begin{equation}
  Q = \left(\begin{array}{cc} Q^{(1)} & Q^{(2+,1)} \\ {[ Q^{(2+,1)} ]}^T & Q^{(2+)} \end{array}\right)\;.
\end{equation}
$Q^{(1)}$ is the submatrix where both $s$ and $t$ are single indices, $Q^{(2+)}$ is the submatrix formed where neither $s$ nor $t$ are single indices, and naturally $Q^{(2+,1)}$ is the case where $u$ is a single index set and $t$ is a multi-index set. We can define by analogy vector forms $C_1=\{C_i\}$ and $C_{2+} = C \setminus C_1$, with associated fluctuations $\delta C_1$ and $\delta C_{2+}$ about the minima of the free energy. The saddle-point equations of the main text derived from the quadratic expansion about the minima (\ref{eq:quadform}) 
 can be written as vector equations
\begin{eqnarray}
  Q^{(1)}\delta C_1  + [Q^{(2+,1)}]^T \delta C_{2+} &=& \beta \delta H \;;\\
  Q^{(2+,1)}\delta C_1  + Q^{(2+)} \delta C_{2+} &=& 0 \label{eq:saddlevector2}\;.
\end{eqnarray}
The second equation can be solved in $\delta C_{2+}$, leaving one equation for $\delta C_1$
\begin{equation}
  Q^{(1)}\delta C_1  - [Q^{(2+,1)}]^T [Q^{(2+)}]^{-1} Q^{(2+,1)} \delta C_{1} =  \beta \delta H \label{eq:step1}\;.
\end{equation}
Since we are expanding about a proper minimum, $Q$ is invertible, and hence so is the submatrix $Q^{(2+)}$. We can identify $\delta C$ with linear responses, it is a sum of perturbations due to each independent field fluctuation
\begin{equation}
  \delta C_s = \beta \sum_z \chi_{s,z} \delta H_z \label{eq:linearresponsevector}\;.
\end{equation}
We denote by $\chi^{NN}$ the $N$ by $N$ matrix with components $d\delta C_i /d \delta H_j$, the off-diagonal elements are sufficient to enforce the correlation constraints $C_\Omega=\{C_{(i,j)}\}$. Equation (\ref{eq:step1}) becomes 
\begin{equation}
  \left[\left(Q^{(1)} \!-\! [Q^{(2+,1)}]^T [Q^{(2+)}]^{-1} Q^{(2+,1)}\right) \!\chi^{NN}\! \right] \delta H \!=\! \delta H\label{eq:step2}\;.
\end{equation}
The result holds for all possible perturbations of the field, so (\ref{eq:step2}) implies
\begin{equation}
  \left(Q^{(1)} - [Q^{(2+,1)}]^T [Q^{(2+)}]^{-1} Q^{(2+,1)}\right) \chi^{NN} = I \;,
\end{equation}
where $I$ is the identity matrix. Finally we separate the coefficient of $\chi^{NN}$ into the energetic part and entropic part ($\Phi$).
Restricting attention to the pairwise model, the energetic term is $-\beta J$ and arises in $Q^{(1)}$, thus the identification between the Hessian $Q$, $\Phi$ and the coupling matrix is 
\begin{equation}
  \Phi - \beta J = \left(Q^{(1)} - [Q^{(2+,1)}]^T [Q^{(2+)}]^{-1} Q^{(2+,1)}\right) \label{eq:generalsol}\;.
\end{equation}
Since we are dealing with a minima the covariance matrix $\chi$ is invertible, we arrive at (\ref{eq:invC}).

In the direct problem we do require knowledge of the structure of higher order response equations. This requires expanding the free energy to 3rd order or higher, which is quite complicated. Aside from maximum entropy, an alternative mechanism for fixing correlations is the use of alternative linear response identities. For example, the three point connected correlation can also be defined
\begin{equation}
  \chi_{s,z} = \frac{\partial^2 F_{CVM}}{\partial H_z \partial \lambda_s} = \frac{\partial C_s}{\partial H_z}\;.
\end{equation}
This quantity can be determined from (\ref{eq:saddlevector2}) and (\ref{eq:linearresponsevector}) as
\begin{equation}
  \chi_{s,z} = \left[[Q^{(2+)}]^{-1} Q^{(2+,1)}\chi^{NN}\right]_{s,z} \;.
\end{equation}
Thus for a three point correlation we might consider $C_{i_1i_2i_3} = (\chi_{i_1i_2,i_3} + \chi_{i_1i_3,i_2} + \chi_{i_2i_3,i_1})/3$ in place of the standard constraint $C_{i_1i_2i_3}=\chi_{i_1,i_2,i_3}$.

\section{Derivation of Sessak-Monasson expression: (\ref{eq:BeliefOffDiagonal}),~\cite{Sessak:SME}} 
\label{app:SessakMonasson}
The form is found from the Bethe approximation to the entropy. 
For the Bethe approximations $Q^{(2+)}$ is a diagonal matrix so the inverse is simple, $Q^{(2+,1)}$ is also sparse
\begin{eqnarray}
  Q^{(2+)}_{ij,ij} &=& \Tr \left[\frac{1}{16 b_{ij}}\right] \nonumber \;;\\
  Q^{(2+,1)}_{ij,k} &=& \Tr \left[\frac{\delta_{k,j}b_i \sigma_i + \delta_{k,i}b_j\sigma_j}{8 b_{ij}}\right]\nonumber\;.
\end{eqnarray}
For a pairwise model
\begin{multline}
  Q^{(1)}_{i,j} = -\beta J_{ij} + \delta_{i,j}\left[\Tr\left[\frac{1}{4 b_{i}}\right] \!\!+\!\!\! \sum_{k(\neq i)} c_{ik}\Tr\left[\frac{b_{k}^2}{4 b_{ik}}\right] \right]  + \\
(1-\delta_{i,j})c_{ij}\Tr \left[\frac{\sigma_i\sigma_j}{4}\left\lbrace \frac{b_{i}b_{j}}{b_{ij}} + \log b_{ij}\right\rbrace \right]\;,
\end{multline}
where $c_{ij}=1 (0)$ for included (excluded) regions. 
Exploiting the sparseness of matrices (\ref{eq:generalsol}) becomes
\begin{equation}
  \Phi^B_{ij} = c_{ij}\Tr \left[\frac{\sigma_i\sigma_j}{4}\left(\frac{b_{i}b_{j}}{b_{ij}} + \log b_{ij}\right) \right] + c_{ij} \frac{Q^{(2+,1)}_{ij,i} Q^{(2+,1)}_{ij,j}}{Q^{(2+)}_{ij,ij}}\;.
\end{equation}
Thus for included regions $c_{ij}=1$, we recover the Sessak-Monasson expression 
\begin{equation}
  [C^{-1}]_{ij} = -J_{ij} + J^{IP}(C) - \frac{C_{ij}}{(1-C_i^2)(1-C_j^2)-C_{ij}^2}\;,
\end{equation}
by combination of (\ref{eq:linresponseidentity}), (\ref{eq:invC}) and (\ref{eq:BeliefOffDiagonal})
, note we needn't know $\lambda$ to estimate $J$ given the correlations.
 
\section{Derivation of $\Delta \Phi^{P_3}$, in the spin-symmetric case: (\ref{eq:DeltaPhiP3})}
\label{app:PhiP3}
If we consider the special case of triangles without symmetry breaking then $C_{s}=0$ for any set of odd parity. We can rearrange the matrices $Q^{(2+)}$ and $Q^{(2+,1)}$, ordering components $Q_{s,t}$ according to the parity, $|s|$ and $|t|$ (notation $|\cdot|$ indicating the number of elements in the set).
Due to symmetry, $Q_{t,u}$ where $|t|$ and $|u|$ are of different parity are null
\begin{eqnarray}
  Q^{(2+)} &=& \left(\begin{array}{cc} Q^{(even)} & 0 \\ 0 & Q^{(odd)} \end{array}\right) \;;\\
  Q^{(2+,1)} &=& \left(0 \qquad Q^{(odd,1)}\right) \;.
\end{eqnarray}
As such the expression (\ref{eq:generalsol}) in a symmetric case becomes
\begin{equation}
  \Phi = (Q^{(1)}+\beta J) - {[Q^{(odd,1)}]}^T [Q^{(odd)}]^{-1} Q^{(odd,1)}\;.
\end{equation}
For the plaquette approximation $Q^{(odd)}$ is block diagonal, and for the case $P_3$ the matrix is diagonal, $Q^{(odd,1)}$ is also sparse. The relevant matrix components are
\begin{eqnarray}
  Q^{P_3}_{ijk,ijk}\!\! &=& \!\!\Tr \!\!\left[\frac{1}{2^6 b_{ijk}}\right]\;;\\
  Q^{P_3}_{ijk,l}\!\! &=& \!\!\Tr \!\!\left[\frac{\delta_{l,i}b_{jk}\sigma_j\sigma_k+\delta_{l,j}b_{ik}\sigma_i \sigma_k + \delta_{l,k}b_{ij}\sigma_i \sigma_j}{2^4 b_{ijk}}\right]\nonumber\;.
\end{eqnarray}
The matrix $Q^{(1)}$ for a pairwise model has components
\begin{multline}
   Q^{P_3}_{i,j} = \delta_{i,j}\left[\sum_{R:i,j\in R} c_R \frac{b_{R\setminus i}^2}{4 b_R}  \right] +  (1-\delta_{i,j})\bigg[-\beta J_{i,j} + \\
      \sum_{R:i,j\in R} c_R\frac{\sigma_i\sigma_j}{4}\left(b_{R\setminus i,j}\log b_R + \frac{b_{R\setminus i}b_{R\setminus j}}{b_R} \right) \bigg]\nonumber\;.
\end{multline}
Again we exploit the sparseness of these matrices to write for the off-diagonal component
\begin{equation}
  \Phi^{P_3}_{i,j} = Q^{P_3}_{i,j} - \sum_{k (\neq i,j)}c_{ijk} \frac{Q^{P_3}_{i,ijk}Q^{P_3}_{j,ijk}}{Q^{P_3}_{ijk,ijk}} \label{eq:PhiT3}\;,
\end{equation}
where for each included plaquette $c_{ijk}=1$ (and zero otherwise). Taking $\Phi^{P_3}-\Phi^B$ (\ref{eq:BeliefOffDiagonal}) and
representing the beliefs in their symmetric forms ($C_i=0,C_{ijk}=0$), we arrive at (\ref{eq:DeltaPhiP3}).

\section{Homogeneous lattice model solutions}
\label{app:HoLatt}
On a translationally invariant lattice we can exploit redundancy of the parameters to find the homogeneous solution. We can label the variables by their geometric coordinates, and we take the standard dot-product on the Euclidean vector space.

A simple Hamiltonian for variables embedded on a hypercubic lattice $\mathcal{L}$ of dimension $d$ is
\begin{equation}
  \mathcal{H} = \sum_{i \in \mathcal{L}} \sum_{\xi \in \Xi}\sigma_i \sigma_{i+\xi}/2 \label{eq:lathomo}\;,
\end{equation}
where $\Xi$ is the set of vectors describing the relative position of coupled variables. Couplings are taken to be $1$ to remove clutter in notation, $\beta$ controls the strength of interaction and we allow $\beta<0$ to describe the antiferromagnetic model. 

For planar lattice models the free energy can be calculated exactly, including in the thermodynamic limit~\cite{Wannier:A,Baxter:ESM}. The energy $E = \sum_{i,j} \langle \sigma_i\sigma_j\rangle$ is related to the mean nearest neighbor correlation ${\hat c}$ by $E=3 {\hat c} N$, forming the basis for the comparison of figure \ref{fig:TrianglularLattice}.

\begin{figure}[!ht]
  \includegraphics[width=0.8\columnwidth]{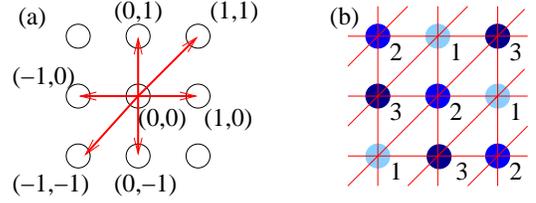}
  \caption{\label{fig:trilat2} (color online) The HTL lattice model embedded on a square lattice. (a) The origin and the relative position of (coupled) nearest neighbors. The elements of the set $\Xi$ that define the displacement of neighbors are shown in red for the HTL. (b) The interactions are tripartite: each sublattice labeled $1$ to $3$ in a regular pattern as shown forms a set of variables that do not self-interact (except possibly at the boundary). The susceptibility of the lattice for $\beta<0$ is largest with respect to fluctuations breaking this symmetry.}
  \vspace{-0.15in}
\end{figure}

In the case of the triangular lattice the nearest neighbors are defined $\Xi=\{\pm (0,1),\pm(1,0),\pm (1,1)\}$ (see figure \ref{fig:trilat2}(a)).

\subsection{Homogeneous symmetric solution}

To avoid clutter we present the simplest case of the spin-symmetric solution. In this simple case the beliefs are determined by a single parameter $C_{i,i+\xi}={\hat c}, \forall \xi \in \Xi$; which describes the solution to (\ref{eq:lathomo}) in the absence of spontaneous symmetry breaking
\begin{eqnarray}
  b^*_i &=& \frac{1}{2} \;;\\\; b^*_{ij} &=& \frac{(1+{\hat c}\sigma_i\sigma_j)}{4} \;;\\
  b^*_{ijk}&=& \frac{1+{\hat c}(\sigma_i\sigma_j+\sigma_i \sigma_k + \sigma_j \sigma_k)}{8}\;.
\end{eqnarray}
Once $\Phi$ is computed ${\hat c}$ is determined as the fixed point of (\ref{eq:linrespiter2}).

For this solution it is also relatively simple to identify symmetry breaking instabilities with respect to homogeneous perturbations of ${\hat c}$, and with respect to perturbations breaking the tripartite symmetry of the HTL (see figure \ref{fig:trilat2}(b)). For finite $L$, where the tripartite symmetry is broken by the boundary conditions (e.g. $L=5$), the full Hessian must be constructed and analysed to determine stability.

By translational invariance $\Phi$ is a function only of the diplacement between its components $\Phi_{i,j}=\phi_{i-j}$.
For a symmetric model the NMF solution is found from $\phi^N_{i} = \delta_{i,0}$.
For the Bethe approximation, with edge regions $\{(i,i+\xi) : \xi \in \Xi\}$ (removing sets that differ only in ordering), the homogeneous solution is $\phi^B_i =\phi^N_i + \Delta \phi^B_{i}$, where
\begin{multline}
  \Delta \phi^B_{i} = \delta_{i,0}\left[\sum_{\xi \in \Xi}\frac{{\hat c}^2}{1-{\hat c}^2}\right] +\\
  \sum_{\xi \in \Xi}\delta_{i,\xi}\left(\atanh({\hat c}) -\frac{{\hat c}}{1-{\hat c}^2}\right)\;.
\end{multline}
The $P_3$ approximation, which has plaquettes $\{(i,i+\xi_1,i+\xi_2): \xi_1+\xi_2 \in \Xi \}$ (without repetitions), is determined by $\phi^{P_3}=\phi^B+\Delta \phi^{P_3}$ with
\begin{multline}
  \Delta \phi^{P_3}_{i} = \delta_{i,0}\left[-\!\!\! \sum_{\xi_1,\xi_2,\xi_3 \in \Xi}\delta_{\xi_1+\xi_2+\xi_3,0}\frac{2 {\hat c}^3}{(1+2{\hat c})(1 - {\hat c}^2)} \right] +\\
 \sum_{\xi_1,\xi_2 \in \Xi} \delta_{\xi_1+\xi_2+i,0} \left[\frac{({\hat c} - {\hat c}^2)^2}{(1 - {\hat c}^2) (1 - 3 {\hat c}^2 + 2 {\hat c}^3)}\right. + \\
 \left. \frac{1}{4}\log\left(1 - \frac{(4 {\hat c}^2)}{(1 + {\hat c})^2}\right)\right]\;.
\end{multline}
To determine the solutions for large $|\beta|$ in practice (\ref{eq:linrespiter2}) is applied making use of damping, annealing and symmetry. 

\subsection{The asymptotic solution}
\label{app:HoLattasy}
Using these approximations we can solve the linear system of equations for lattice models
\begin{equation}
  (-\beta J + \Phi({\hat c})) \chi = I\;.
\end{equation}
For a translationally invariant lattice embedded in a hyper-cubic lattice of dimension $d$,  we can exploit the Fourier representation of $\chi_{i,j}$
\begin{equation}
  \chi_{i,j} = \chi_{i-j} = \int d \mu {\tilde \chi}(\mu) \exp(- \rmi 2 \pi (i-j)\cdot \mu)\;.
\end{equation}
with the integral over the unit hyper-cube centered on the origin. The inverse transform is
\begin{equation}
  {\tilde \chi}(\mu) = \sum_z \chi_{z} \exp(\rmi 2\pi \mu \cdot z) \;.
\end{equation}
A general solution in the thermodynamic limit ($L\rightarrow \infty$) is found for (\ref{eq:lathomo}) as
\begin{equation}
  {\tilde \chi}(\mu) = \frac{1}{\phi_0 + \sum_{\xi \in \Xi} [\phi_\xi-\beta] \exp(2 \pi \rmi \mu \cdot \xi)}\label{eq:gensolF}\;.
\end{equation}
From which any element of the matrix $\chi_{a,b}$ can be constructed by the inverse discrete Fourier transform.
In the standard framework ($\lambda_{nn}=0$), we solve for ${\hat c}$ by minimization, and calculate once $\chi$. For the new method we begin with an estimate of ${\hat c}^0$ and update according to (\ref{eq:linrespiter2}), for the special case of the triangular homogeneous lattice (\ref{eq:gensolF}) yields
\begin{widetext} 
\begin{equation}
  {\hat c}^{t+1} = \frac{1}{3} \int_{-1/2}^{1/2} d \mu_1 \int_{-1/2}^{1/2}  d \mu_2  \frac{\cos(2 \pi \mu_1) + \cos(2 \pi \mu_2) + \cos(2 \pi (\mu_1+\mu_2))}{\phi_0({\hat c}^t) + 2[\phi_1({\hat c}^t)-\beta] [\cos(2 \pi \mu_1) + \cos(2 \pi \mu_2) + \cos(2 \pi (\mu_2+\mu_2))]}\;.
\end{equation} 
\end{widetext}
where owing to homogeneity $\phi_\xi=\phi_1$ and
\begin{equation}
  \phi_i = \delta_{i,0}\phi_0 + \sum_{\xi \in \Xi} \delta_{i,\xi} \phi_1\;.
\end{equation}

\section{High temperature expansion}
\label{sec:hightempexpansion}
Loops cause the failure of the Bethe approximation, and the shortest loops contribute the leading order errors in $\beta$. To understand errors up to $O(\beta^X)$ it is sufficient to consider a diagrammatic expansion, where diagrams of size greater than $X$ do not contribute. The free energy can be explicitely constructed, as such the exact connected correlations and entropy, $\Phi_{i,j}$ and $L_i$, this allows a comparison to our approximation which can again be determined by a diagramatic expansion.
The diagrams contributed to the leading order corrections in Bethe are apparent on a single triangle (fully connected graph of 3 variables), whilst for the $P_3$ approximation it is sufficient to consider a tetrahedron (fully connected graph of 4 variables). The two smallest graphs that are not solved exactly by the respective approximations.

Regarding the inverse problem, in the new method $\chi$ and $C$ coincide with the data, whereas in the standard method $\chi$ and $C_1$ coincide with the data and $C_{2+}$ is determined by maximum entropy. The couplings $J_{ij}$ is determined by (\ref{eq:invC}), hence the error arises only in the term $\Phi_{i<j}(C)$. The fields $H_i$ are determined by (\ref{eq:saddleM}), so the error arises in $\sum_j J_{ij} C_j + L_i$. We call $C^*_{2+}$ the value determined by maximum entropy in the standard approach, as a perturbation about the data ($C$) we can solve the linearized saddle-point equation to determine the leading order error
\begin{equation}
  0 = \frac{\partial F}{\partial C_{2+}}(C) + [Q^{(2+,2+)}(C)] (C_{2+}^*-C_{2+}) \;.
\end{equation} 
The sampling error in the data can dominate the method error in practical scenarios for both methods, for high fidelity data and sufficiently large $\beta$ it is the errors outlined that are most significant. Our analysis assumes the sampling error is negligible compared to the error in the entropy approximation. 

In the direct problem the error on $C$ depends in a coupled manner upon the errors in all $\Phi$, $L$ and the maximum entropy procedure for $C\setminus C_\Omega$. Thus the increased accuracy of $\Phi_{i<j}$ is not necessarily realised in increased accuracy of $C_{ij}$ for example. A minor modification, including the on-diagonal constraints discussed in Appendix \ref{app:ondiagoffdiag}, can ameliorate this error~\cite{Raymond:Correcting}. 
 
We demonstrate the leading order diagrams for the saddle-point term $L_i$ and the matrix $\Phi$ for the weak coupling limit ($J$ small,$\beta$ finite), using notation $\doteq$ to indicate the asymptotic nature. For the weak coupling limit we abbreviate $t_i=\tanh(\beta H_i)$ and $T_i=1-t_i^2$.

\subsection{Bethe, errors on fully connected model}
The error in inference of $H$ is limited by
\begin{equation}
  L^B_i - L^{exact}_i \doteq 2 \beta^3  t_i \sum_{j < k (\neq i)}T_jT_k J_{ij} J_{ik} J_{jk} \;,\label{eq:errorLi}
\end{equation}
errors of $O(J^3,\beta^4)$ respectively in the weak coupling and high temperature expansions.
The error in inference of $J$ is limited by, for $i\neq j$,
\begin{multline}
\!\! \Phi^B_{ij} - \Phi^{exact}_{ij}\!\! \doteq \!\! - 2  \beta^4 \sum_{k (\neq i,j)} T_k (2 J_{kij} + 2 J_{jki} + J_{ijk}) J_{ik} J_{jk} \label{eq:errorplaq}\;,
\end{multline}
where $J_{ijk}=J_{ik} J_{jk} t_i t_jT_k$, errors of $O(J^4,\beta^5)$ are found. The high temperature result is dominated by a different set of diagrams, as shown (\ref{eq:beta5}).
The on-diagonal component error in $\Phi$ is determined by the diagrams
\begin{equation}
  \Phi^B_{ii} - \Phi^{exact}_{ii} \doteq 2 \beta^3 \sum_{j<k (\neq i)}T_jT_k J_{ij} J_{ik} J_{jk}\;,\label{eq:errorPhii}
\end{equation}
respectively $O(J^3,\beta^3)$. This error is not significant for the inverse problem, only the direct problem.

If we fix correlation parameters according to maximum entropy, the standard method, new errors are introduced
\begin{equation}
  C^*_{ij}-C_{ij} \doteq - \beta^2 T_iT_j\sum_{k\neq (i,j)} T_k J_{i,k}J_{j,k} \;,
\end{equation}
which are $O(J^2,\beta^2)$.
The error on $\Phi_{i<j}$ for exact data (\ref{eq:errorplaq}) is worsened after considering this additional error source to $O(J^4,\beta^4)$, thus in the high temperature limit we gain one order magnitude in the inference of $J_{ij}$. In the weak coupling limit fewer diagrams contribute to the error but the order remains the same (unless $H=0$). The diagrams contributing to the leading order errors in $\Phi_{ii}$ (\ref{eq:errorPhii}) and $L_i$ (\ref{eq:errorLi}) are the same after considering this error source, but contribute with opposite sign. Overestimations become underestimations and vice-versa, a pattern we see realised in both direct and inverse problem applications even for models with stronger coupling. 

\subsection{Plaquette, $P_3$ errors on fully connected model}

Defining $J_{ijkl}=J_{ij}J_{jk}J_{kl}J_{il}T_jT_kT_l$ we can write 
\begin{multline}
   L^{P_3}_i - L^{exact}_i \doteq - 2 \beta^4 t_i \sum_{j<k<l} [J_{ijkl}+J_{ikjl}+J_{ijlk}]\;, \label{eq:Lplaq}
\end{multline}
the error is $O(J^4,\beta^5)$. For the off-diagonal component
\begin{multline}
   \Delta \Phi^{P_3}_{ij} \doteq - 2 \beta^4 J_{ij}t_i t_j \sum_{k<l (\neq i,j)}  J_{kl} [J_{ik}J_{jl} + J_{il}J_{jk}] T_k T_l\;,\label{eq:PhiOffDiagplaq}
\end{multline}
the error is $O(J^4,\beta^6)$. For the special case of a strictly pairwise Hamiltonian $H=0$ we recover (\ref{eq:beta7}) for the new method, the leading order diagram is $O(\beta^7)$. For the on-diagonal component, 
\begin{equation}
  \Delta \Phi^{P_3}_{ii} \doteq -2 \beta^4\!\!\sum_{j<k<l (\neq i)}\![J_{ijkl}+J_{ikjl}+J_{ijlk}]\;, \label{eq:PhiDiagplaq}
\end{equation}
the errors are $O(J^4,\beta^4)$. 

If we fix correlation parameters according to maximum entropy, the standard method, new errors are introduced
\begin{eqnarray}
 \!\!\! C^*_{ij}\!-\!C_{ij} \!\!&\doteq&\!\! \beta^3 T_iT_j\!\!\!\!\!\!\sum_{k<l (\neq i,j)}\!\!\!\!T_kT_lJ_{kl}(J_{il} J_{jk} + J_{ik} J_{jl})\;;  \\
 \!\!\! C^*_{ijk}\!-\!C_{ijk} \!\!&\doteq&\!\! 2 \beta^3 T_iT_jT_k\sum_{l (\neq i,j,k)} t_l T_l J_{il}J_{jl}J_{kl}\;.
\end{eqnarray}
As for the Bethe approximation the forms (\ref{eq:Lplaq}) and (\ref{eq:PhiDiagplaq}) undergo a sign change with this modification. In the case of high temperature the off-diagonal component error in $\Phi$ is worsened relative to the case of exact correlations $C$ (\ref{eq:PhiOffDiagplaq}), to $O(J^4,\beta^5)$. In the high temperature limit we thus improve the estimation of $J_{ij}$ by an order of magnitude, in the special case of zero external field the improvement is two orders of magnitude.


\section{Diagonal and on-diagonal constraints}
\label{app:ondiagoffdiag}
In this paper we have argued that since the CVM approximation parameters should be properly interpreted as marginal probabilities it is sensible to consider the most accurate beliefs possible, which may be linear response estimates rather than maximum entropy estimates. We have then shown how the the free energy should be modified to make the linear response self-consistent. 

An alternative argument could be that for any variational framework two different derivatives (responses), or functions of derivatives, that determine the same quantity ought to be in agreement for a good approximation, those at lowest order being most important. Consider for example the two possible estimates of the pair correlation $E[\sigma_i \sigma_j]$ which can be estimated either by a first derivative with respect to $J_{ij}$, or by a non-linear function of the responses 
\begin{equation}
  \frac{\partial F_{CVM}}{\partial J_{ij}} = \frac{\partial^2F_{CVM}}{\partial H_i\partial H_j}  + \left[\frac{\partial F_{CVM}}{\partial H_{i}}\right]\left[\frac{\partial F_{CVM}}{\partial H_{j}}\right] \label{eq:offdiagonal}\;.
\end{equation}
This constraint is equivalent to (\ref{eq:linresponseidentity}) restricted to the case of pairs. Another important relation at second order in the derivatives would be the self-response
\begin{equation}
  1 - \left[\frac{\partial F_{CVM}}{\partial H_{i}}\right]^2 = \frac{\partial^2 F_{CVM}}{\partial H_i^2} \label{eq:ondiagonal}\;,
\end{equation}
which is an identity not considered in this paper, but has been proposed as a constraint simultaneously in two recent papers~\cite{Raymond:Correcting,Yasuda:SPDC}. The power of the on-diagonal constraint is readily apparent, at the NMF level approximation one already recovers the adaptive TAP equations which are well tested~\cite{Huang:adaTAP,Opper:TAPMpre,Opper:TAPMprl}.

It is natural to call the first class of constraints (\ref{eq:offdiagonal},\ref{eq:linresponseidentity}) off-diagonal, and the second class (\ref{eq:ondiagonal}) on-diagonal, since they relate to the simplest possible off and on-diagonal identities for the response matrix $\chi$ at second order. We have shown in this paper the off-diagonal constraints can yield performance gains in isolation, combining both constraints can also be effective~\cite{Raymond:Correcting}.

\section{CVM free energy with effective fields and couplings}
\label{app:Free_energy_modified_fields}
The constraint that pair connected correlation parameters are fixed leads to a simple modification of the entropy in (\ref{eq:lamC}). In the framework of beliefs the additional terms are non-linear functions of the beliefs, and if $\Omega$ contains only pair of variables (pair constraints) the same entropy can be written concisely in terms of the beliefs
\begin{multline}
  S_\lambda(b) = S(b) - \\\sum_{(i,j) \in \Omega}\;\; \sum_{R: (i,j)\in R} c_R \left(\Tr[b_R \sigma_i \sigma_j] - \Tr[b_R \sigma_i]\Tr[b_R \sigma_j] \right)\;.
\end{multline}
It is clear that this new term is neither linear, convex nor concave in the beliefs, which makes certain standard and robust methods of minimization for the CVM method defunct~\cite{yedidia,Yuille:cccpalgorithms}. Recall $c_R$ are the counting numbers for region $R$ in the approximation, and the sum restricted to inclusion of any subset (e.g. $\{R: i \in R\}$ or $\{R:i,j \in R\}$) sums to one.

If we define a new set of variational parameters, with corresponding constraints
\begin{equation}
  M_i = \sum_{R: i \in R} c_R \Tr [b_R \sigma_i] \label{eq:Mdef}\;,
\end{equation}
we can redefine the entropy as 
\begin{multline}
  S_\lambda(b,M) = S(b) - \sum_{(i,j) \in \Omega}\;\; \sum_{R: (i,j)\in R} c_R \biggr(\Tr[b_R \sigma_i \sigma_j] -  \\
  \Tr[b_R \sigma_i]M_j - \Tr[b_R \sigma_j]M_i + M_i M_j\biggr)\;.
\end{multline}
At which point we notice a convenient factorization for the full free energy
\begin{multline}
  F_{CVM}(J,H,b,M) =\sum_i {\tilde H}_i \Tr[b_i \sigma_i] + \\\sum_{ij}{\tilde J}_{ij} \Tr[b_{ij} \sigma_i \sigma_j] - \frac{1}{\beta} S(b) \label{eq:Fcvm2}\;,
\end{multline}
which is the same as a standard CVM free energy but with effective fields and couplings
\begin{eqnarray}
  {\tilde H}_i &=& H_i + \frac{1}{\beta} \sum_R c_R \sum_{j \in R} M_j \lambda_{ij}\;;\\
  {\tilde J}_{ij} &=& J_{ij} - \frac{1}{\beta} \lambda_{ij} \label{eq:effective_field}\;,
\end{eqnarray}
and additional variational parameters $\{M_i\}$ in one to one correspondence additional constraint (\ref{eq:Mdef}). In the Bethe case the effective field simplifies to
\begin{equation}
  {\tilde H}^B_i = H_i + \sum_{j \in \partial_i}\lambda_{ij}M_j\;,
\end{equation}
where $\partial_i$ are the neighbors of $i$.

If $M$ and $\lambda$ were fixed external parameters, our method would be equivalent to a reassignment of $J$ and $H$. However, since $M$ is variational there is a new reaction term in the saddle-point equation for $M$, and a new linear response term (even for paramagnetic solutions, where $M=0$), each linear in $\lambda$. Parameters $\lambda$ are fixed according to this modified linear response criteria. As such intuition can be counter intuitive. For example by decreasing $\lambda$ we increase the effective coupling, but the intuition that the susceptibility towards a ferromagnetic solution would increase is incorrect, infact the susceptibility of the solution decreases.

By converting the free energy to the form (\ref{eq:Fcvm2}) it becomes clearer how to apply standard methods to construct message passing framework alternatives to the iterative approach we outline, and what may cause instabilities of these frameworks. In particular we note that for fixed $M$ and $\lambda$ we can use any standard message passing or susceptibility propagation procedure~\cite{yedidia}, the subtlety is then in the selection of update rules for $\lambda$ and $M$ (which can be chosen to require only local information).

\end{document}